\documentclass{article}

\usepackage{arxiv}

\usepackage[utf8]{inputenc} 
\usepackage[T1]{fontenc}    
\usepackage{hyperref}       
\usepackage{url}            
\usepackage{booktabs}       
\usepackage{amsfonts}       
\usepackage{nicefrac}       
\usepackage{microtype}      
\usepackage{lipsum}

\usepackage[version=3]{mhchem}
\usepackage{gensymb}
\usepackage{graphicx}
\usepackage{float}

\title{A C/\ce{V2O5} core-sheath nanofibrous cathode with mixed-ion intercalation for aluminium-ion batteries}

\author{
  Nicolò Canever, Thomas Nann\thanks{Corresponding author}\\
  School of Mathematical and Physical Sciences\\
  The University of Newcastle\\
  Callaghan, NSW 2308\\
  Australia\\
  \texttt{thomas.nann@newcastle.edu.au} \\
}

\begin{document}
\maketitle

\begin{abstract}
A new nanofibrous material, consisting of a conductive carbon core and an external layer made of vanadium oxide, has been studied as a cathode for aluminium-ion batteries. The material enables a mixed-ion intercalation mechanism, resulting in the alternating insertion of $\text{Al}^{3+}$ and $\text{AlCl}_4^-$ in the $\text{V}_2\text{O}_5$ and carbon layers, respectively. This is a highly desirable feature for cathode materials which may increase the energy density of future batteries by optimising the utilisation of the electrolyte.
\end{abstract}

\begin{figure*}[h!]
    \centering
    \includegraphics[width=9cm]{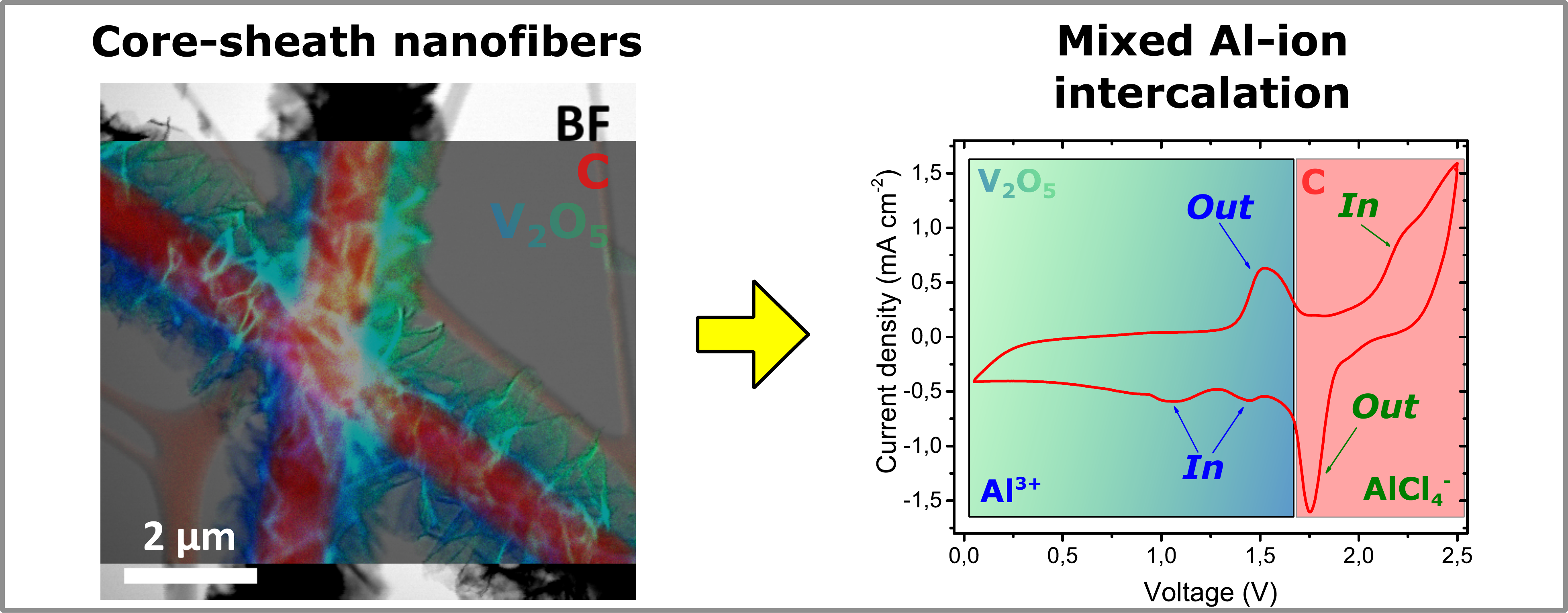}
\end{figure*}


\setcounter{footnote}{0}
The recent rise in the worldwide adoption of renewable energy vectors caused an increased demand for large-scale electrochemical energy storage. Current battery technologies such as lithium-ion, however, often lack the cost-effectiveness and safety requirements necessary for large-scale, grid-level energy storage applications. Therefore, it is important to search for alternative technologies, which are more suitable for this purpose. Non-aqueous aluminium-ion batteries (AIBs) are a promising emerging battery technology with several advantages over existing technologies and promising performance.\cite{ambroz_trends_2017,k.das_aluminium-ion_2017,canever_acetamide:_2018,munoz-torrero_critical_2019} One of the biggest challenges of these systems is maximising the energy density of the devices: this is due to the fact that the electrolyte takes part in the electrochemical reactions within the battery,\cite{kravchyk_efficient_2017} in a similar fashion to lead-acid cells. A recent publication described the possibility of using an organic compound as the cathode material for AIBs, employing an energy storage mechanism that involves the complexation of \ce{AlCl2+} ions. By forming a composite with graphite, which is known to involve the intercalation of \ce{AlCl4-} ions, \cite{lin_ultrafast_2015,wang_advanced_2017} the cathode allowed a mixed-ion mechanism, which the authors argued would allow a higher energy density by exploiting the species present in the ionic liquid electrolyte more efficiently.\cite{kim_rechargeable_2018}

In this work, we present a material that aims at similar properties, and exploits the synergistic effects of carbon and vanadium oxide to enable a mixed-ion intercalation mechanism. Both materials are well established in the AIB literature, and have been shown to work individually as intercalation-based cathodes: graphitic carbon allows for the insertion of tetrachloroaluminate (\ce{AlCl4-}) ions upon charging, while \ce{V2O5} has been shown to facilitate the insertion of \ce{Al^3+} on discharge.\cite{jayaprakash_rechargeable_2011,gu_confirming_2017} Therefore, by combining the two materials in one cathode, a mixed-ion mechanism could take place, allowing for an increased energy density. In order to achieve such a material, a nanofibrous core/sheath structure was prepared by adaption of a method first described by Li \textit{et al.},\cite{li_flexible_2015} in which a solvothermal synthesis was used to grow nanostructured \ce{V2O5} on a carbon nanofibre (CNF) substrate. The preparation method for this material was relatively straightforward: first, CNFs were synthesised through a well-known procedure involving the electrospinning and carbonisation of polyacrylonitrile\cite{nataraj_polyacrylonitrile-based_2012}, then a solvothermal process was performed on the fibres, followed by thermal annealing, to grow \ce{V2O5} nanosheets on the surface of the CNFs.

\begin{figure*}[ht]
\centering

\begin{minipage}{0.31\textwidth}
	\includegraphics[width=\linewidth]{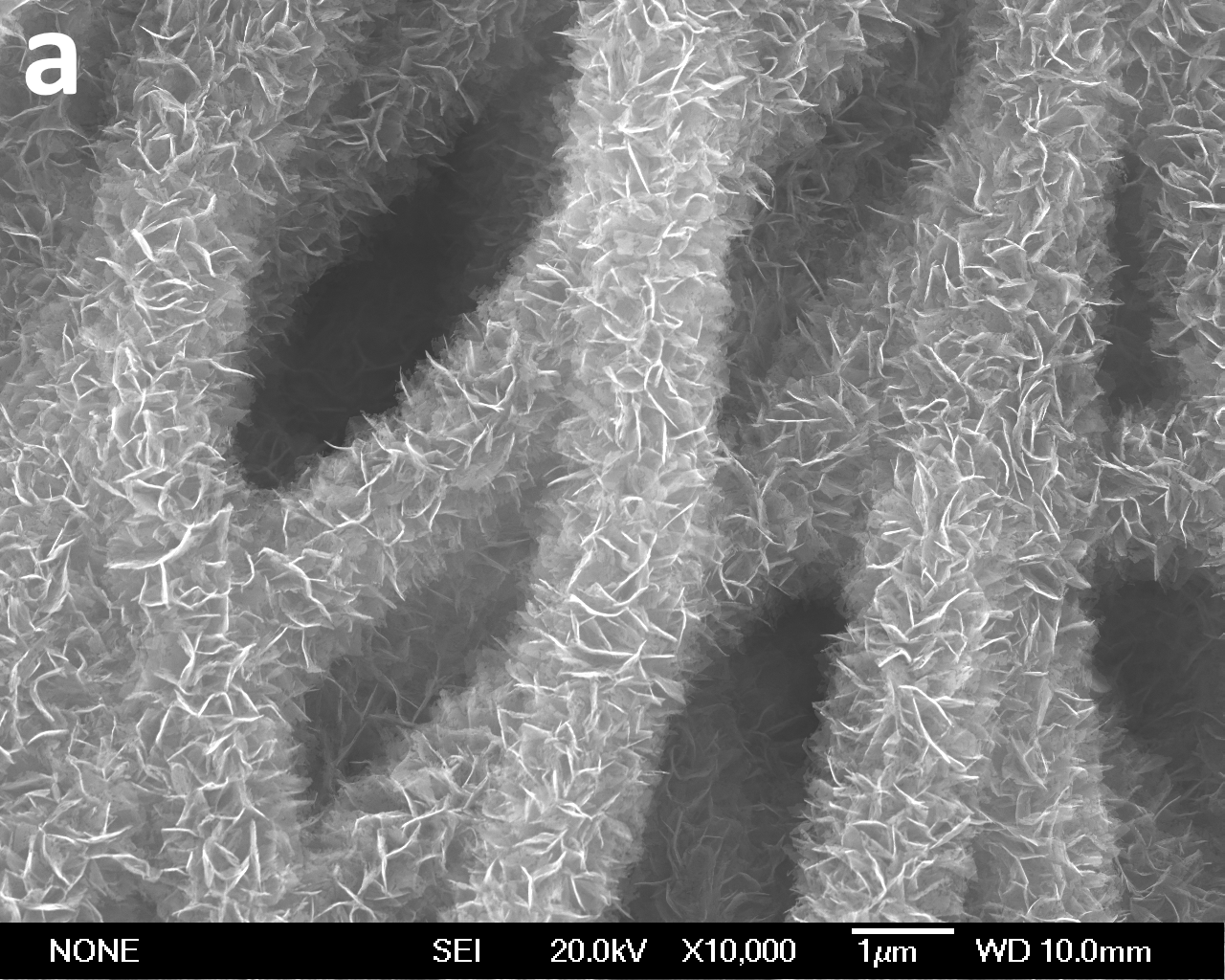}
\end{minipage}
\hspace{0.3cm}
\begin{minipage}{0.51\textwidth}
	\vspace{0.3cm}
	\includegraphics[width=\linewidth]{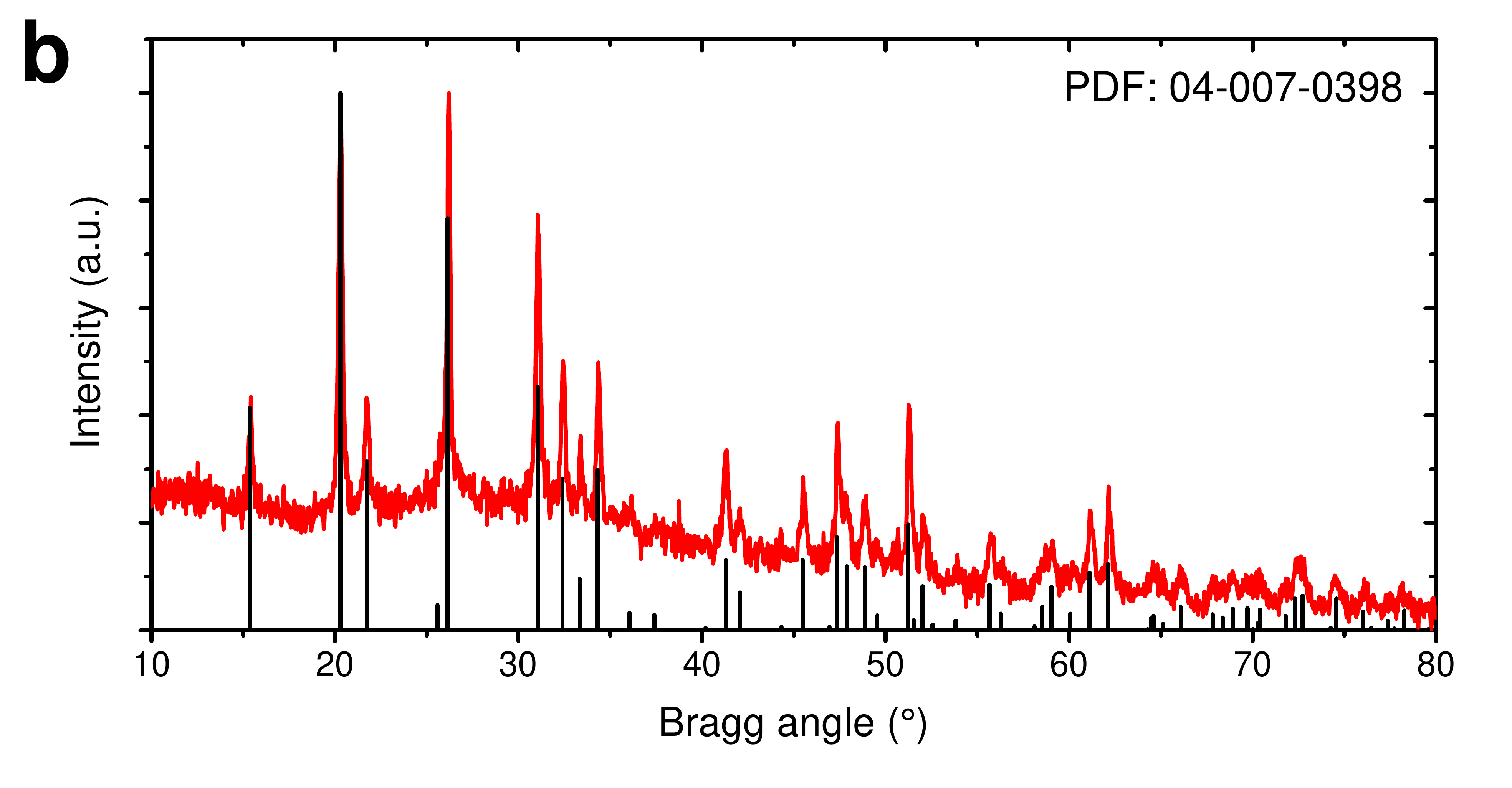}
\end{minipage}

\par
\vspace{0.1cm}

\begin{minipage}{0.35\textwidth}
 \centering
 \includegraphics[width=0.49\linewidth]{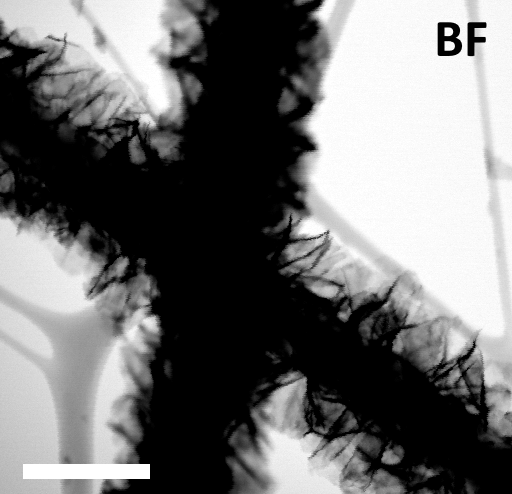}
 \includegraphics[width=0.49\linewidth]{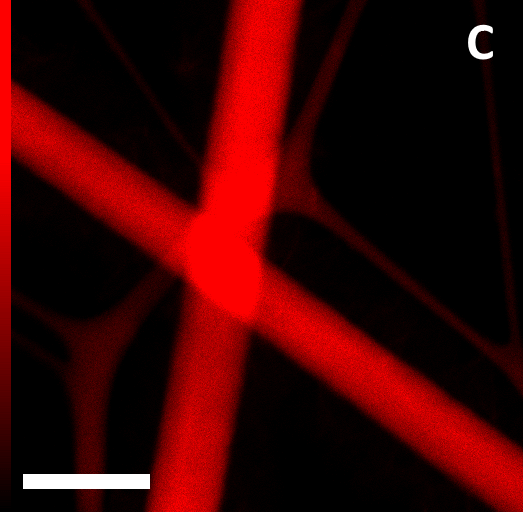}
 \includegraphics[width=0.49\linewidth]{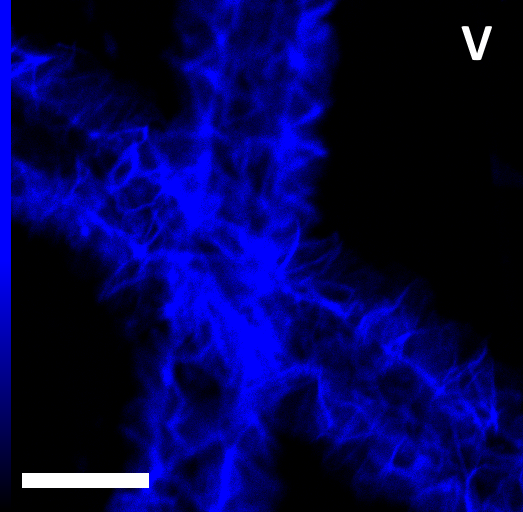}
 \includegraphics[width=0.49\linewidth]{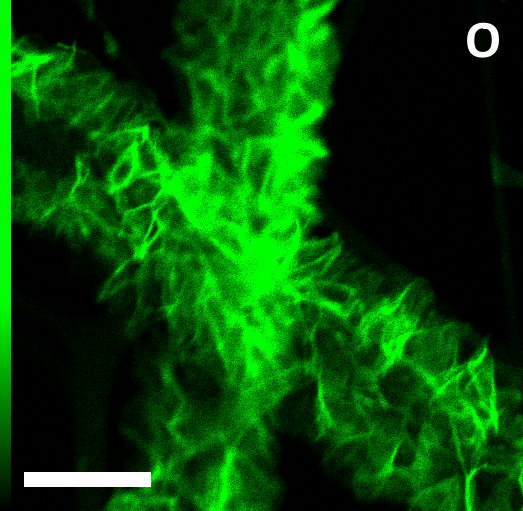}
 \end{minipage}%
\begin{minipage}{0.35\textwidth}
 \centering
 \includegraphics[width=0.965\linewidth]{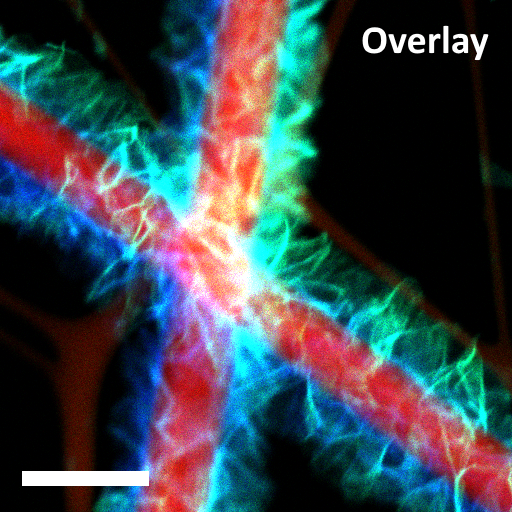}
\end{minipage}
 \caption[STEM-EDX maps of vacuum-annealed solvothermal core-sheath NF]{Top level: (a) SEM micrograph and (b) X-ray diffractogram of the core/sheath C/\ce{V2O5} nanofibres, overlaid with the matching database file (ICDD powder diffraction file (PDF): 04-007-0398) for \ce{V2O5}.  Bottom level: STEM-EDX elemental maps of the core/sheath C/\ce{V2O5} nanofibres: STEM-BF image (top left), carbon (red), vanadium (blue), oxygen (green) maps, and overlay image (right). Scale bar of STEM image and maps: 2 $\mu$m.}
\label{fig:coresheathEMXRD}
 \end{figure*}

Scanning electron microscopy (SEM) imaging (Fig.\  \ref{fig:coresheathEMXRD}a) and scanning transmission electron microscopy energy disperse X-ray (STEM-EDX) elemental mapping (Fig.\  \ref{fig:coresheathEMXRD} - bottom level) have been performed on the material. The images revealed that the material featured a nanofibrous structure with \ce{V2O5} nanosheets --- a few tens of manometers in thickness --- protruding from the fibres. Furthermore, it can be seen from the STEM-EDX elemental maps that the carbon core was successfully retained in the material, even after the annealing step. X-ray diffractometry was also conducted on the sample (Fig.\ \ref{fig:coresheathEMXRD}b): a series of peaks in good agreement with one of the International Centre for Diffraction Data (ICDD) database patterns for orthorombic \ce{V2O5} ($\alpha$-\ce{V2O5}) can be seen in the diffractogram, indicating that the correct phase of vanadium oxide was successfully obtained in the process. 

In order to test the performance of the material in a battery device, Swagelok-type cells were assembled using the core/sheath nanofibres (NFs) as the cathode. Due to the presence of a carbon core, the material exhibited good mechanical properties, which allowed the nanofibre mats to be used as "self standing" electrodes. The material was directly cut into appropriately shaped discs, which were used directly as cathodes in the assembly of the devices without further treatments or additives such as binders or conductivity enhancers. High purity aluminium foil was used as the negative electrode, and a 1.3:1 molar mixture of aluminium trichloride (\ce{AlCl3}) and 1-ethyl-3-methylimidazolium chloride ([EMIm]Cl) was used as the electrolyte. A galvanostatic charge-discharge test was initially performed on the devices using a current rate of 100 mA g$^{-1}$, employing a "discharge first" protocol. 
\begin{figure}
	\centering
	\includegraphics[width=0.45\linewidth]{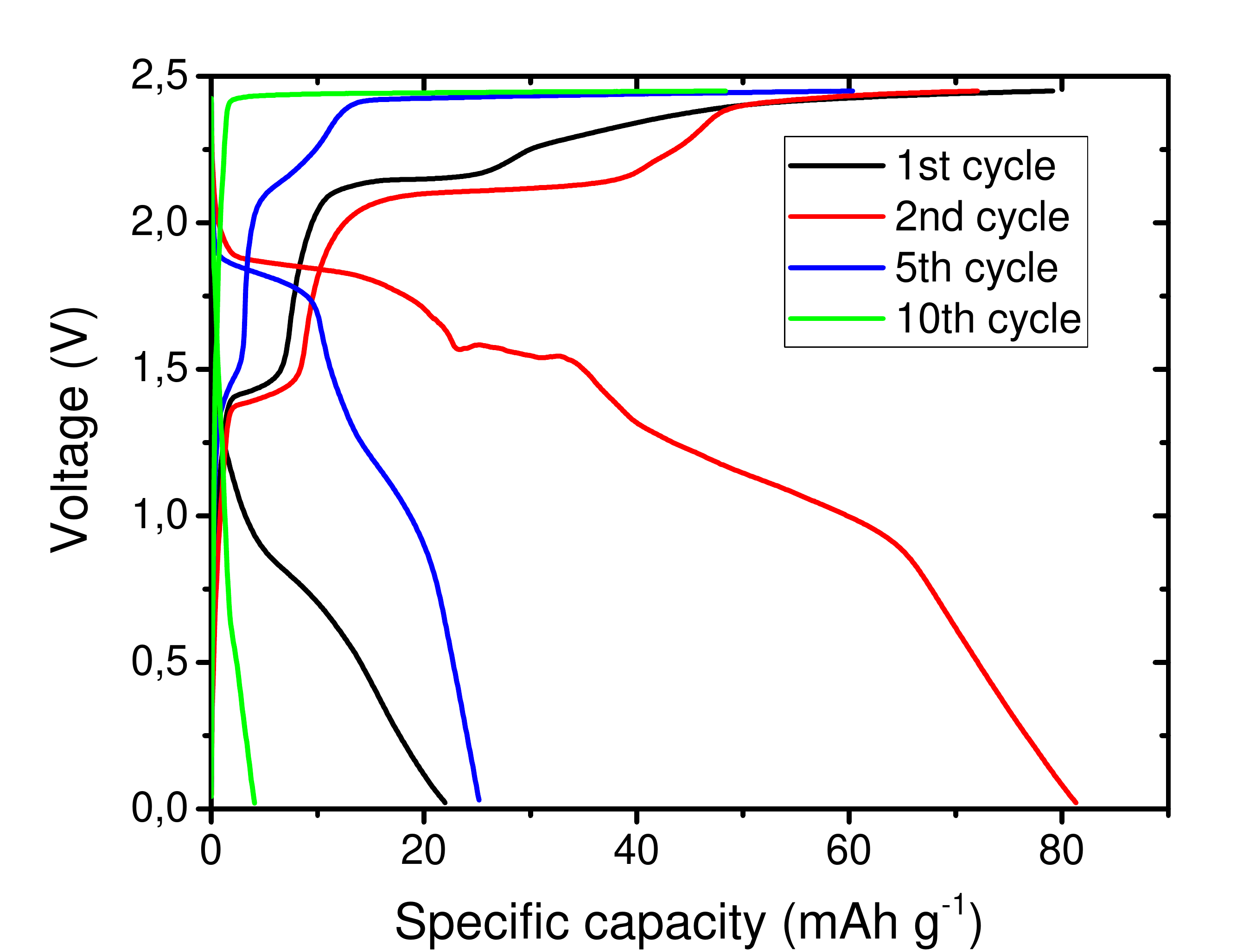}

	\caption[Galvanostatic cycling of core/sheath C/\ce{V2O5} NF]{Galvanostatic charge-discharge  profiles of a Swagelok-type cell built using core/sheath C/\ce{V2O5} nanofibres as cathode. }
	\label{fig:CoreSheathv240DC}
\end{figure}

\begin{figure*}[t]
	\centering
	\includegraphics[width=0.9\textwidth]{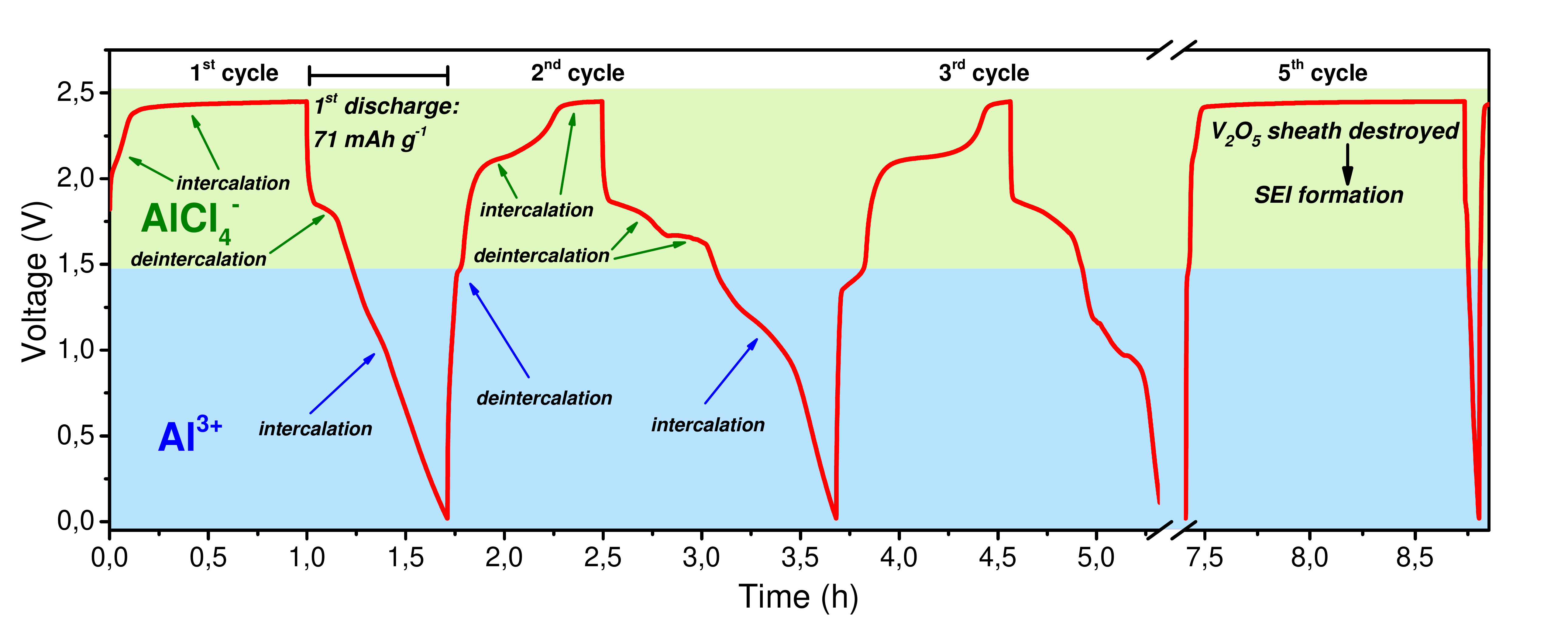}
	\caption[Galvanostatic cycling of core/sheath C/\ce{V2O5} NF (charge first)]{Galvanostatic cycling (at a cycling rate of 100 mA g$^{-1}$), starting with a charge sequence, of a Swagelok-type cell built using core/sheath C/\ce{V2O5} nanofibres as cathode.}
	\label{fig:CoreSheathChargefirst}
\end{figure*}

An interesting behaviour was observed in this experiment: it can be seen from Fig.\ \ref{fig:CoreSheathv240DC} that the first discharge cycle featured only a weakly defined plateau at about 0.8 V and yielded an overall modest capacity of about 21 mAh g$^{-1}$. The following charging sequence, however, showed a series of sharp features: a short plateau at about 1.4 V and two longer ones around 2.1 V and above. The specific capacity relative to the first charge was also considerably longer, at about 80 mAh g$^{-1}$. The following discharge sequence then contained yet more new plateaus that were not present in the first cycle, at 1.8, 1.6, and 1.2 V. The specific capacity yielded was also notably higher, at roughly 81 mAh g$^{-1}$. This behaviour strongly differs from the typical galvanostatic charge-discharge profiles of \ce{V2O5} cathode materials in AIBs, in which the presence of weakly defined plateaus below 1 V was usually observed. \cite{jayaprakash_rechargeable_2011,gu_confirming_2017,wang_binder-free_2015,chiku_amorphous_2015} In the following cycles, the features in the charge-discharge profiles became more regular. However, the capacity faded quite rapidly and stabilised around 4 mAh g$^{-1}$ after about 10--11 cycles (see also ESI Figure S1).

A few conclusions can be drawn from the behaviour observed in the galvanostatic charge-discharge test: the presence of a much longer charge sequence than the corresponding discharge in the first cycle, accompanied by the appearance of new features, indicates the possibility of multiple processes taking place in the material. These processes appear to be happening within different voltage ranges: one process below 1.5 V and the other between 1.5 and 2.45 V. We believe that the material undergoes a mixed-ion intercalation mechanism, involving both \ce{Al^3+} and \ce{AlCl4-} ions:

\begin{itemize}
 \item When the battery operates at lower voltages (roughly below 1.5 V), the primary mechanism during the discharge is the expected insertion of \ce{Al^3+} ions into \ce{V2O5}, as demonstrated by previous reports. \cite{gu_confirming_2017} The valence of the vanadium ion is therefore expected to go from \ce{V^5+} in the charged state, to a mixture of \ce{V^4+} and \ce{V^3+} in the discharged state. \cite{chiku_amorphous_2015,wang_new_2013}
 
 \item At higher voltages, the galvanostatic profile resembles the one observed for graphitic materials.\cite{kravchyk_efficient_2017,lin_ultrafast_2015,wang_advanced_2017} This would suggest that at higher voltages, a similar process would take place in the nanofibres: the intercalation of \ce{AlCl4-} ions during the charging phase and their deintercalation during the discharge phase.
\end{itemize}

The intercalation of \ce{AlCl4-} ions has likely been taking place in the carbon core of the NFs. Our previous work has shown that low-graphitic CNFs do not normally allow an intercalation-based energy storage, due to the presence of surface defects causing the formation of a solid-electrolyte interphase (SEI).\cite{canever_solid-electrolyte_2020} On the other hand, if the surface of the CNFs was adequately shielded from direct contact with the electrolyte, the SEI formation process was inhibited, and a certain degree of intercalation could take place in the material. 
It can be seen from Fig.\ \ref{fig:CoreSheathv240DC} that during the first cycles, the typical asymptotic charging voltage plateau behaviour, correlated with SEI formation (as discussed in our previous publication),\cite{canever_solid-electrolyte_2020} was not observed. In this case, the \ce{V2O5} sheath layer could therefore serve a dual purpose: an active material for the intercalation of \ce{Al^3+} ions and a protective layer for the surface defects of CNFs, enabling the mixed-ion process. 
This interpretation is also supported by the fact that after the first few cycles, the discharge capacity decreased to very low values and an asymptotic voltage plateau appeared in the charging phases: this change in electrochemical behaviour could have been caused by the disruption of the \ce{V2O5} sheath, leading to the termination of the shielding effect. Consequently, this would revert the energy storage mechanism to the one observed in pure, non-coated CNFs.\cite{canever_solid-electrolyte_2020}

In order to further test the hypothesis of a multiple-ion intercalation mechanism, a second galvanostatic charge-discharge test, using the same voltage range and cycling rate, was performed. This time, however, a protocol starting with a charging step was used: this way, if the \ce{AlCl4-} intercalation were to take place, then a charging plateau should be detectable in the first charging step, even without performing any previous discharging steps. It can be seen from Fig.\  \ref{fig:CoreSheathChargefirst} that two charging plateaus were indeed observed in the first cycle, which are therefore likely associated with the insertion of \ce{AlCl4-}. Furthermore, the first discharge sequence yielded a discharge capacity of 71 mAh g$^{-1}$, a value much higher than the one obtained in the test starting with the discharge step (Fig.\ \ref{fig:CoreSheathv240DC}). This result indicates that a certain degree of charging has taken place in the first step, which adds up to the overall discharge capacity observed in the first cycle. This test then suggested that when core/sheath C/\ce{V2O5} nanofibres are used as a cathode in AIBs, the freshly assembled battery device would start in an intermediate state of charge. Therefore either a charging or a discharging step could be first performed in the cycling protocol. Although this test alone does not provide conclusive proof, it indicates that a mixed-ion mechanism is definitely possible:

\begin{itemize}
 \item If the first step performed on the device is a charging phase, then the first event taking place at the cathode is the intercalation of \ce{AlCl4-}. In the following discharge step, the deintercalation of \ce{AlCl4-} then takes place from 2.45 to 1.5 V and the insertion of \ce{Al^3+} takes place as the potential decreases below 1.5 V. Finally, \ce{Al^3+} is deintercalated in the following charge cycle, completing the cycle. This scenario is shown in Fig.\  \ref{fig:CoreSheathChargefirst}.
 \item If instead the first step performed on the device is a discharging phase, then the first event taking place at the cathode is the intercalation of \ce{Al^3+}. The ions are then deintercalated in the following charge step until the 1.5 V potential is reached, after which the intercalation of \ce{AlCl4-} takes place. In the next discharge step, \ce{AlCl4-} are deintercalated from 2.45 to 1.5 V, completing the cycle.
\end{itemize}

Further evidence for this hypothesis was given by the high discharge capacity values obtained in the first few cycles.
A maximum discharge capacity of about 118 mAh g$^{-1}$, corresponding to a remarkable specific energy of about 156 mWh g$^{-1}$, was obtained in the second cycle (see also ESI, Fig.\ S1). Such high values suggest that both the core and the sheath components of the cathode material are likely contributing to the energy storage mechanism. Assuming that both components could individually undergo their respective mechanisms, this would imply that both processes could be happening at the same time in a core-sheath NF: intercalation of \ce{AlCl4-} ions into carbon during the charging phase and insertion of \ce{Al^3+} into \ce{V2O5} during the discharging phase. This is a very interesting feature, as the intercalation of multiple different ionic species correlates to a more efficient utilisation of the chloroaluminate electrolyte.\cite{kim_rechargeable_2018,liang_halfway_2018}

In both the galvanostatic cycling experiments the performance is unfortunately quite unstable, as discharge capacities fade rather quickly. A possible explanation for this could be that factors such as the mechanical stress induced by the intercalation and deintercalation reactions, or the gradual dissolution of the material in the electrolyte, could cause the gradual destruction of the \ce{V2O5} sheath. This would cause a loss in discharge capacity, and disrupt the synergistic interaction between the core and sheath materials that allowed the multiple-ion mechanism to take place. This was also supported by the change in shape of the galvanostatic profiles (Fig.\  \ref{fig:CoreSheathChargefirst}): after the first two or three cycles, the features and plateaus of the charging and discharging sequences started becoming gradually less pronounced, suggesting the disintegration of the cathode. This hypothesis also fits with the assumption regarding the role of the \ce{V2O5} sheath in inhibiting SEI formation; the removal of the sheath layer would mean that the surface of CNFs was again exposed to the electrolyte, causing the same type of behaviour explained in our previous work.\cite{canever_solid-electrolyte_2020} Indeed, it can be seen that the tenth and fifth cycles in Fig.\  \ref{fig:CoreSheathv240DC} and \ref{fig:CoreSheathChargefirst}, respectively, closely resembled the galvanostatic profiles of devices built using self-standing, low graphitic carbonaceous materials, featuring a predominantly capacitive behaviour: a very long plateau was observed near the upper voltage cutoff that once again suggests the formation of an SEI, and is paired with an extremely short discharging sequence without any outstanding features, indicating a predominantly capacitive behaviour. \cite{canever_solid-electrolyte_2020,wang_kish_2017,stadie_zeolite-templated_2017} Further evidence for the disintegration of the sheath layer was found in scanning electron microscopy and cyclic voltammetry data (See also ESI, Fig.\ S2 and S3).

\begin{figure*}
\centering

 \begin{minipage}{0.35\textwidth}
 \centering
 \includegraphics[width=0.49\linewidth]{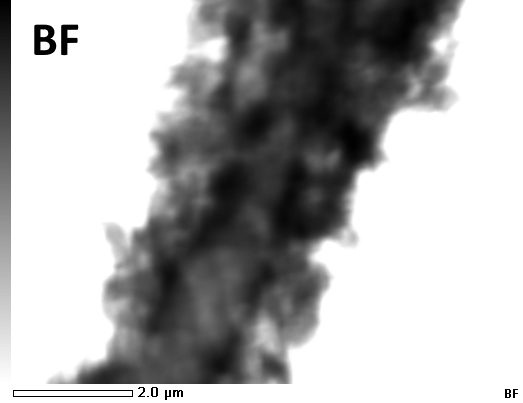}
 \includegraphics[width=0.49\linewidth]{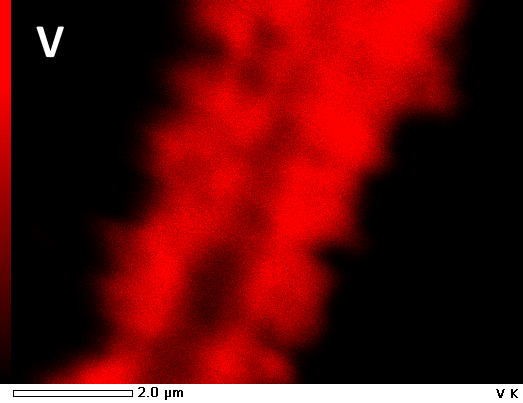}
 \includegraphics[width=0.49\linewidth]{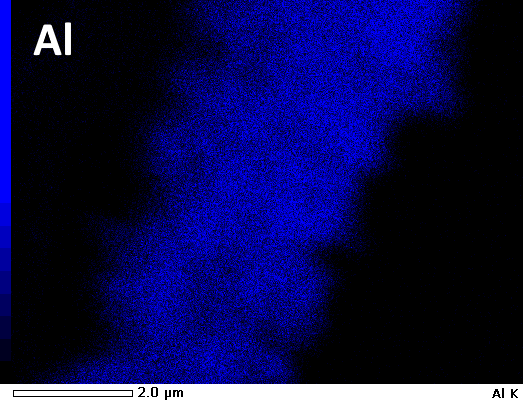}
 \includegraphics[width=0.49\linewidth]{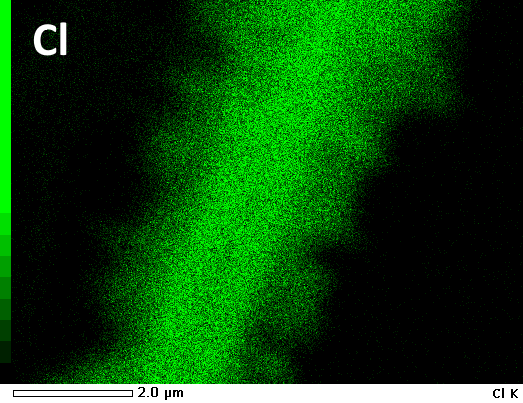}
 \end{minipage}%
\begin{minipage}{0.347\textwidth}
 \centering
 \includegraphics[width=0.98\linewidth]{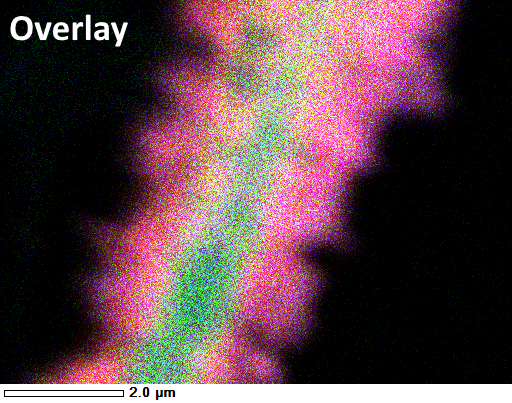}
\end{minipage}

\par
\vspace{0.1cm}
\begin{minipage}{0.42\textwidth}
\includegraphics[width=0.9\linewidth]{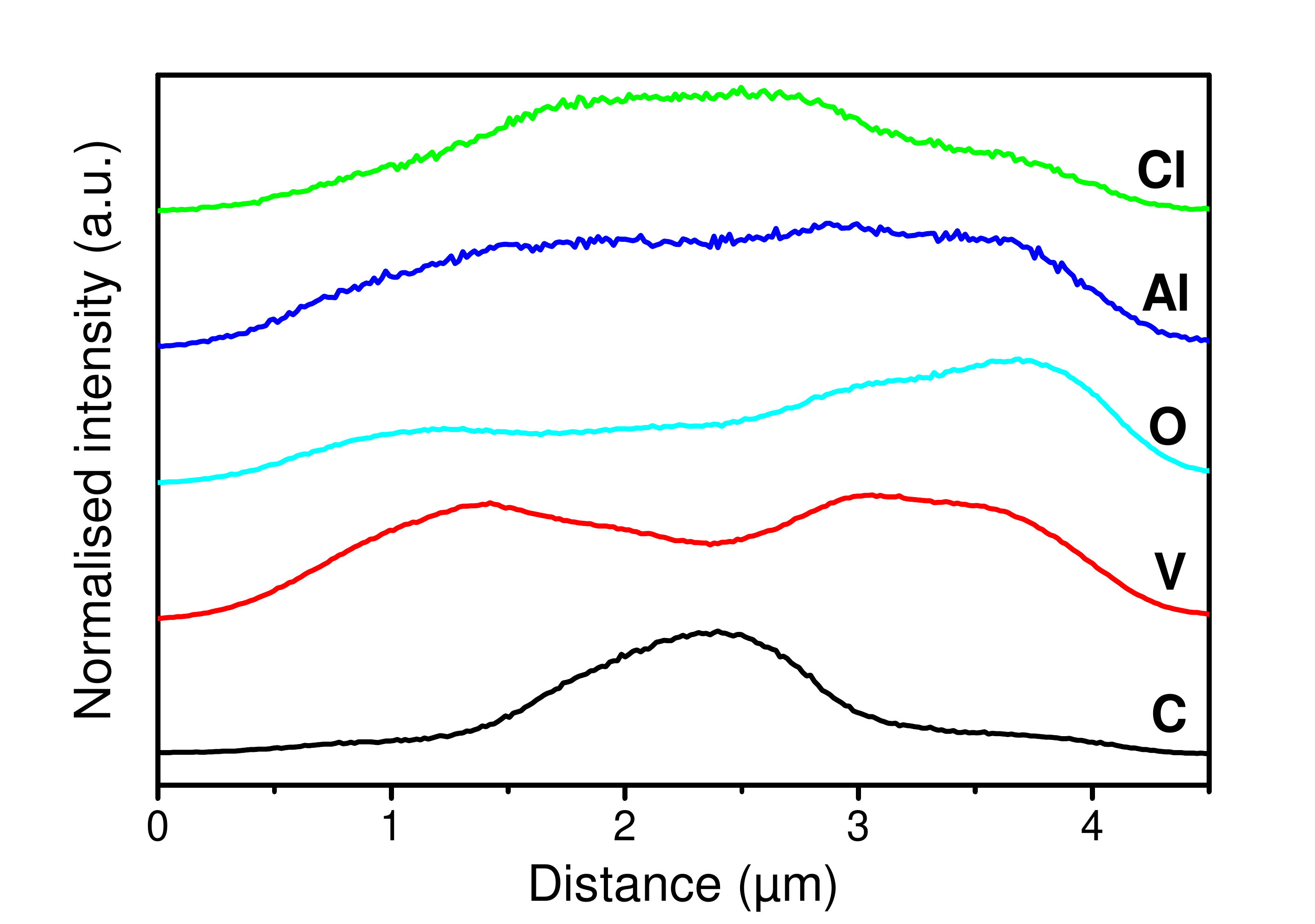}
\end{minipage}
\begin{minipage}{0.32\textwidth}
\includegraphics[width=0.9\linewidth]{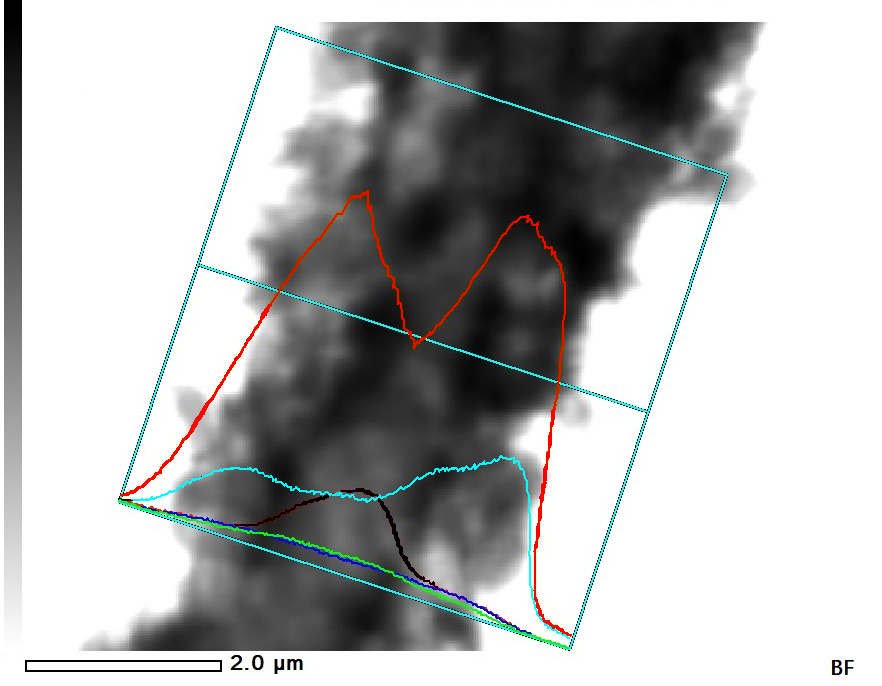}
\end{minipage}
\caption[ \textit{Ex situ} STEM-EDX maps of core/sheath C/\ce{V2O5} NF]{ \textit{Ex situ} STEM-EDX data of charged-state core/sheath C/\ce{V2O5} NF. (Top level) Elemental maps: STEM-BF image (top left), vanadium (red), aluminium (blue), chlorine (green) maps, and overlay image (right). (Bottom level) Linear intensity profiles of individual elements across a single fibre (left), and corresponding integration area (right).}
\label{fig:CoreSheathExSituTEM}
 \end{figure*}
 
In order to investigate the energy storage process and verify the presence of a mixed-ion mechanism in the material, an \textit{ex-situ} TEM analysis was performed (Fig.\ \ref{fig:CoreSheathExSituTEM}). First, a device using the core/sheath C/\ce{V2O5} NFs as cathode was assembled. Then, a galvanostatic charging protocol consisting of a charge, a discharge, and a charge (i.e.\ one and a half cycles) at a current rate of 100 mA g$^{-1}$ was performed on the battery. Finally, the cell was disassembled and the cathode was observed using TEM and STEM-EDX mapping, to assess the aluminium and chlorine content in the fibres. 
There are a few reasons for choosing this specific charging-discharging protocol for the experiment: first of all, imaging the cathode after a limited number of cycles is preferable, due to the probability for the sheath layer to disintegrate. Furthermore, the previous tests (Figs.\  \ref{fig:CoreSheathv240DC} and \ref{fig:CoreSheathChargefirst}) have shown that the largest specific capacities were obtained in the second cycle of the device. Therefore, by disassembling the device in the second cycle, the largest perturbation of the system would be observed, which would ideally correspond to the most evident change in elemental composition of the material. 
Finally, our experiments suggested that if the intercalation of \ce{AlCl4-} were happening at the cathode, it would have to take place during the charging step. Therefore, by disassembling the device and observing the cathode in its charged state, detecting the presence of both Al and Cl in the carbon core by STEM-EDX mapping would be a possible confirmation that this process indeed takes place.

It can be observed from Fig.\ \ref{fig:CoreSheathExSituTEM} that both aluminium and chlorine are present in the nanofibres after the charging step.
Furthermore, it can be seen that the intensity of the Cl K$_\alpha$ peak is notably more pronounced in the core section of the fibres, where C is present. On the other hand, its intensity is visibly lower in the \ce{V2O5} sheath. This can be interpreted as evidence that the \ce{AlCl4-} ion can effectively intercalate into the carbon core in a similar fashion to graphitic cathode materials, whereas \ce{Al^3+} is inserted into the \ce{V2O5} sheath, which is coherent with the previous reports.\cite{gu_confirming_2017} It is unclear at this stage how the \ce{AlCl4-} ions are diffusing past the sheath layer to intercalate into the carbon core. Further investigation would be required to determine the exact mechanism. 
It is also worth noting that the intensity of the Al peak, on the other hand, is mostly uniform across the entire fibre(s). 
According to the proposed intercalation mechanism, the intensity of Al should be lower than the one observed in the sample, which is in a charged state. However, the high concentration of Al in the sheath layer can be justified by the entrapment of \ce{Al^3+} ions in the \ce{V2O5} layer due to the poor reversibility of the intercalation-deintercalation processes. This is consistent with the results observed for \ce{V2O5} cathodes in earlier literature reports.\cite{jayaprakash_rechargeable_2011,gu_confirming_2017,wang_binder-free_2015}

\begin{figure}
	\centering
	\includegraphics[width=0.45\linewidth]{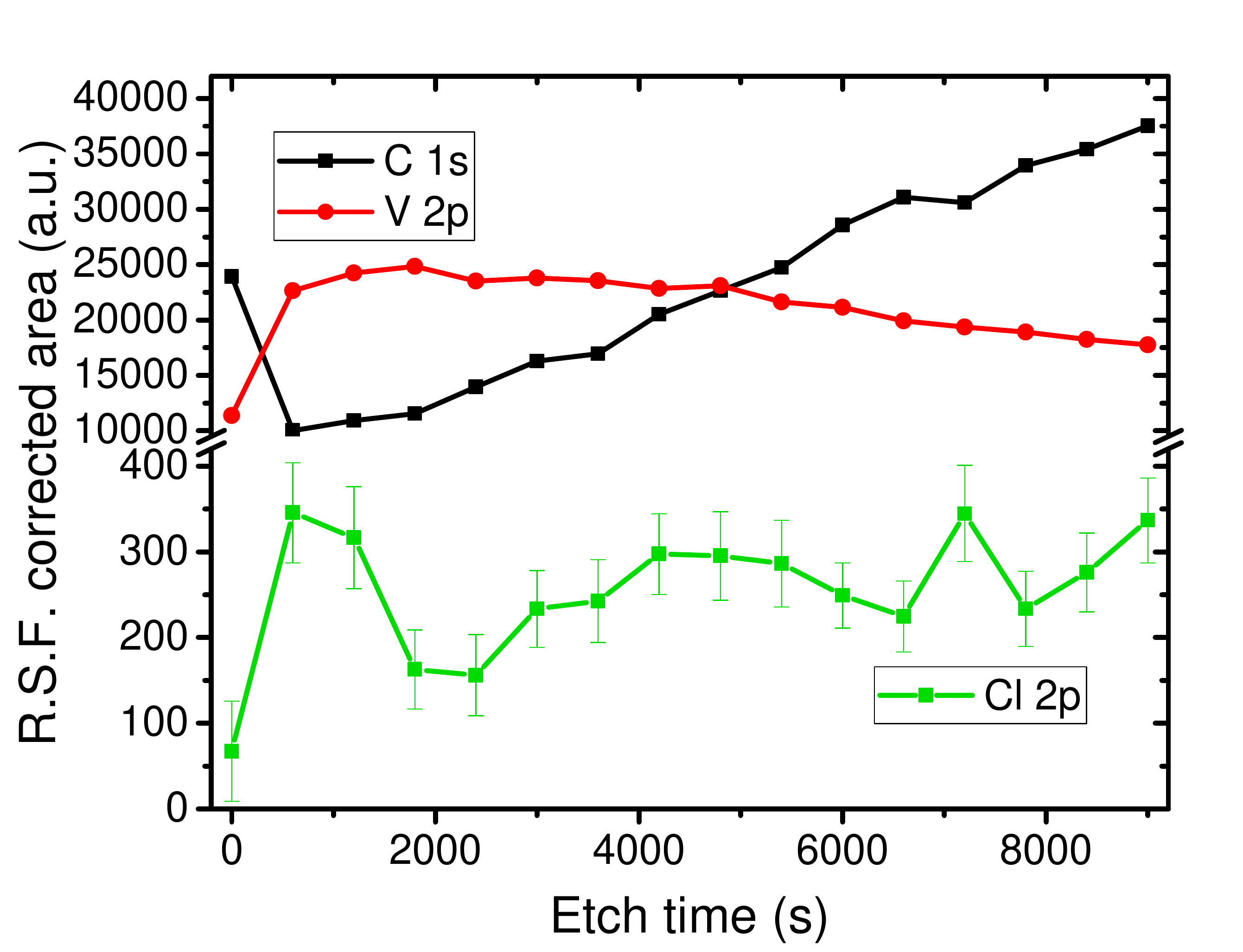}
	\caption[ \textit{Ex-situ} XPS depth profile of core-sheath C/\ce{V2O5} NF]{ \textit{Ex-situ} XPS depth profile of charged-state core/sheath C/\ce{V2O5} NFs. The error bars on the Cl 2p data points refer to the standard deviation calculated on the integration area by the CasaXPS software.}
	\label{fig:CoreSheathdepthprofile}
\end{figure}

A depth profile study using X-ray photoelectron spectroscopy (XPS) was also performed on the same cathode as the \textit{ex-situ} TEM study.
While the previous STEM-EDX elemental mapping analysis focused on a single fibre, the use of this technique should provide a more averaged, "bulk"-type information of the elemental composition of the sample. This was because the instrument scans across a larger area compared to that used in the previous TEM experiment, therefore sampling a larger quantity of material. It is also worth noting that although the thickness of the "sheath" layer is several hundreds of nm, a large amount of it is composed of empty space. Furthermore, based on the diameter of the nanofibres observed in the previous TEM study, it would appear that part of the \ce{V2O5} layer was lost in the charge-discharge process. Therefore, the validity of this technique in identifying the elemental composition of the "core" layer of the nanofibres is reasonable and worth attempting. 

The depth profile data is displayed in Fig.\  \ref{fig:CoreSheathdepthprofile}. It can be seen from the plot that the integrated areas for the C 1s and V 2p peaks follow an ascending and descending trend, respectively. This is expected given the morphology of the sample. Furthermore, by correcting the values according to their relative strength factor (RSF), the intensity of the carbon peak actually surpasses that of the vanadium peak after an etching time of approximately 4800 seconds, suggesting that the core level of the nanofibres could have possibly been reached by the prolonged etching cycles. This assumption is however mostly speculative, as the data obtained from the XPS spectra reflects the composition of an ensemble of nanofibres illuminated by the X-ray source. Furthermore, other factors such as the integration range of the peaks in the spectra could also be influencing the intensities of the peaks, leading to an overestimation or underestimation of the elemental composition. Possibly the most important information that can be observed from the depth profile is the trend of the Cl 2p peak: similar to carbon, that the area of the Cl 2p peak also follows a roughly increasing trend. Although the data points feature a significant deviation, this can be seen as further evidence that chlorine is present in the carbon core of the fibres, as a consequence of the intercalation of \ce{AlCl4-} ions during the charging step.\footnote{Unfortunately, due to a combination of the low sensitivity of the XPS technique for aluminium\cite{moulder_handbook_1993} and the overlap between its most intense (2p) peak and the 3s peak of vanadium, a reasonably accurate depth profile for this element could not be obtained.} This result therefore further supports the hypothesis of a multi-ion intercalation mechanism taking place in the material. It is also worth noting that the first three data points relative to the Cl 2p depth profile follow a seemingly contrasting trend compared to the remaining data. Although many factors could be causing these outliers, one possible explanation for this is that a certain amount of electrolyte or SEI layer could still be present on the nanofibres, resulting in a high concentration of chlorine-containing species in the outermost layers of the sample. This is also partially supported by the presence of a trend inversion in the carbon and vanadium concentration for the first data point as well, which could be attributed to the presence of [EMIm]$^+$ ions or other organic species deriving from the decomposition of the electrolyte.

In summary, core/sheath C/\ce{V2O5} nanofibres were successfully fabricated and their performance in AIBs have been tested. Experimental evidence suggests that a mixed-ion mechanism, involving the "alternating" intercalation of both \ce{Al^3+} and \ce{AlCl4-}, is taking place in the material. This is made possible by the \ce{V2O5} layer inhibiting the SEI formation process that would normally happen in CNF cathodes (Figure S4) by limiting the contact between the surface defects of CNFs and the electrolyte. Because of this, the carbon core can allow the insertion of \ce{AlCl4-} ions in the same way as graphitic materials do.
This is a highly desirable feature, as this mechanism would potentially maximise the energy density of the system by optimising the utilisation of the electrolyte.\cite{kim_rechargeable_2018,liang_halfway_2018} It is worth noting that this strategy would be only effective if the electrolyte mass used in the device is high enough to be the primary contributor to the overall mass of the device, as a larger cathode mass would be also required.

The synergistic behaviour of the core and sheath layers is unfortunately short-lived, as the \ce{V2O5} sheath tends to disintegrate after a few cycles, exposing the carbon core and leading to the formation of an SEI and overall poor performance. Further work could be performed to find a material with a more stable behaviour, which could allow this mixed-ion intercalation mechanism over a large number of cycles.

\section*{Acknowledgements}

We would like to thank David Flynn for his support with the electron microscopy facilities at Victoria University of Wellington. Additional thanks to Dr.\ Colyn Doyle from the University of Auckland for running the X-ray photoelectron spectroscopy measurements featured in this work.

\section*{Data availability}
The data that support the findings of this study are available upon request from the authors.

\bibliographystyle{ieeetr}  
\bibliography{refs}  

\end{document}


\beginsupplement
\title{\textbf{A C/\ce{V2O5} core-sheath nanofibrous cathode with mixed-ion intercalation for aluminium-ion batteries}}
\author{Nicolò Canever and Thomas Nann}
\date{}
\maketitle

\section*{Electronic Supplementary Information (ESI)}

\subsection*{Experimental section}
\sloppy Polyacrylonitrile (PAN), N,N'-dimethylformamide (DMF), vanadium oxytriisopropoxide (VOTIP), isopropanol, 1-ethyl-3-methylimidazolium chloride ([EMIm]Cl) and aluminium trichloride were purchased from Sigma-Aldrich and used without any additional treatment, unless differently specified. Al foil was purchased from MTI Corp.

SEM micrographs were acquired with a JEOL-JSM6500F field emission scanning electron microscope, using an accelerating voltage of 20 kV. 

TEM micrographs and STEM-EDX elemental maps were acquired with a JEOL-JEM2100F transmission electron microscope, using an accelerating voltage of 200 kV. Samples were prepared by dispersing the material in ethanol via ultrasonication and drop casting onto a lacey carbon grid.

For \textit{ex situ} analysis, the cycled devices were disassembled inside the glovebox, and the cathodes were washed with ethanol to remove any residual electrolyte. 
X-ray diffractograms were acquired using a PANalytical X'pert PRO XRD diffractometer operating in reflection mode, using a Cu K$\alpha$ X-ray source, in the Bragg angle range between 10 and 80 degrees. Samples were prepared by lightly grinding the materials and depositing them on a zero-background holder.

XPS spectra were acquired using a Kratos Axis UltraDLD X-ray photoelectron spectrometer (Kratos Analytical, Manchester, UK) equipped with a hemispherical electron energy analyser. Spectra were excited using monochromatic Al K$_\alpha$ X-rays (1486.69 eV) with the X-ray source operating at 150 W. This instrument illuminates a large area on the surface and then using hybrid magnetic and electrostatic lenses collects photoelectrons from a desired location on the surface. 
The depth profile used a mini-beam III (MBIII) ion gun. A 5 kV \ce{Ar+} ion beam was rastered over a 3mm x 3mm area. To avoid etch crater wall effects, the depth profile data was collected using the FOV1 lens and the slot aperture approx. 220 $\mu$m$^2$. The analysis chamber was at pressures in the 10$^{-9}$ torr range throughout the data collection. Data analysis was performed using the CasaXPS software. The peak area was calculated by direct integration, without the use of curve fitting.

Cyclic voltammetry (CV) tests were performed using a Metrohm Autolab PGSTAT128N potentiostat, operating in a 2-electrode configuration: the cathode terminals of the devices were connected to the working electrode, and the anode terminals were connected to the counter and reference electrode. The current density is reported against the macroscopic surface area of the electrode, and does not account for the nanostructure of the electrode.

\sloppy Galvanostatic charge-discharge experiments were performed using a NEWARE BTS CT-4008-5V10mA-164 (MTI Corp.) battery analyser system. The specific capacities are referred to the total cathode mass.

\subsubsection*{Fabrication of core/sheath C/\ce{V2O5} nanofibres}

A precursor solution was prepared by adding 0.8 mL of PAN (average M$_w$: 150,000) to 10 mL of DMF under magnetic stirring. The solution was then loaded into a plastic syringe with a stainless steel, 18-gauge, blunt tip needle. The electrospinning process was then performed, using a voltage of 15 kV, a flow rate of 1 mL/hour, and a tip-to-collector distance of 13 cm. A rotating drum collector was used in the process, to optimise the form factor of the nanofibre mats.
The resulting fibres mats were peeled from the current collector, and then a first stabilisation process was performed by heating in air, using a muffle furnace with the following heating programme: heating at 5 \degree C/minute until 220 \degree C, holding the temperature for 2 hours, and cooling naturally to room temperature. Finally, carbon nanofibres (CNFs) were obtained by carbonising the material under a nitrogen atmosphere, using a tube furnace with the following heating programme: heating at 4 \degree C/minute until 200 \degree C, hold temperature for 20 minutes, heating at 10 \degree C/minute until 600 \degree C, hold temperature for 20 minutes, heating at 5 \degree C/minute until 1400 \degree C, hold temperature for 1 hour, cooling at 5 \degree C/minute until 200 \degree C, and finally cooling naturally to room temperature.

Following the literature procedure by Li \textit{et al.},\cite{li_flexible_2015} a precursor solution was prepared by stirring 200 $\mu$L of VOTIP into 30 mL of isopropanol. The solution was then transferred into a stainless steel autoclave with a 50 mL teflon-lined vessel, and a 3 cm $\times$ 3 cm CNF sheet was then added. The autoclave was then sealed and heated at 200 \degree C for 10 hours. Then, the nanofibre sheet was removed from the solution, washed three times with water and ethanol, and dried in air. Finally, the nanofibres were annealed by heating them at 240 \degree C overnight in a vacuum oven. 

\subsubsection*{Electrolyte preparation}

[EMIm]Cl was baked at 100 \degree C in a vacuum oven for 48 hours to remove its water content, and immediately brought inside a nitrogen-filled glovebox with the \ce{O2} and \ce{H2O} levels kept below 1 ppm.
The electrolyte was prepared by gradually adding 1.3 equivalents of anhydrous \ce{AlCl3} to 1 equivalent [EMIm]Cl inside the aforementioned glovebox. An exothermic reaction takes place, after which a brown, slightly viscous liquid is formed. 

\subsubsection*{Device construction}

\begin{figure}[h!]
    \centering
    \includegraphics[width=0.3\textwidth]{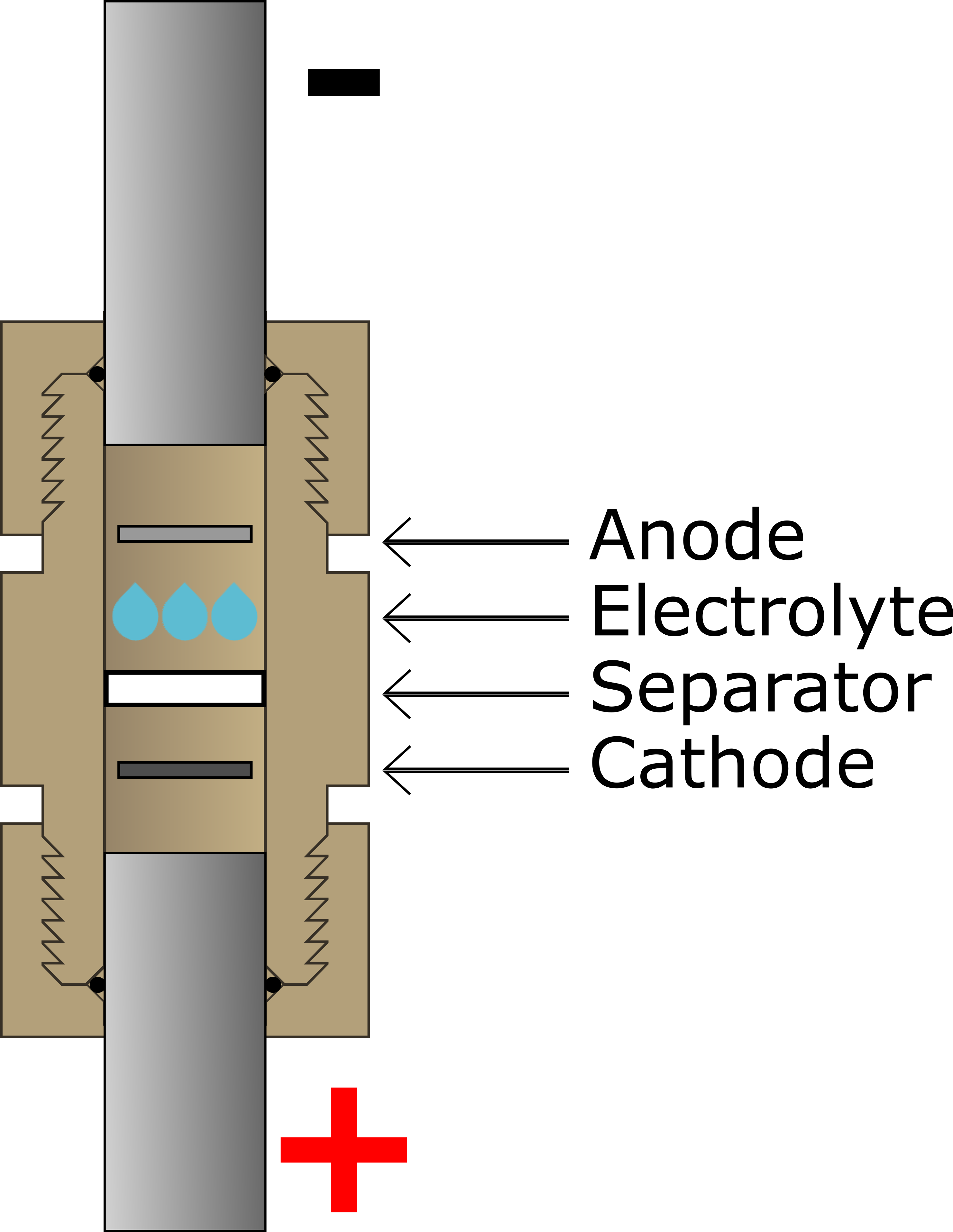}
    \caption{Schematic representation of the custom-built Swagelok-type cell used for the electrochemical tests shown in this work.}
    \label{Swagelok}
\end{figure}

Nanofibre mats were cut into 11 mm diameter discs to be used directly as cathode in the devices. Prototype batteries were then assembled using a custom-built Swagelok-type cell consisting of a cylindrical PEEK casing with an inner diameter of 12 mm and two Mo rods as the current collectors (Fig. \ref{Swagelok}). All the components of the device were first baked at 100 \degree C in a vacuum oven for at least 2 hours to remove any residual water, then immediately transferred inside a nitrogen-filled glovebox with the \ce{O2} and \ce{H2O} levels kept below 1 ppm. One of the previously prepared NF discs was used as a cathode, and 11 mm diameter, high purity Al foil discs were used as the anode. Glass microfibre (GF/D) discs with 12 mm diameters, soaked in approximately 300 $\mu$L of the electrolyte, were used as separators. The devices were then wrapped with Parafilm as an additional moisture barrier, and taken outside the glovebox for electrochemical testing.

\subsection*{Supplementary data}

\begin{figure}[h!]
	\centering
	\includegraphics[width=0.49\textwidth]{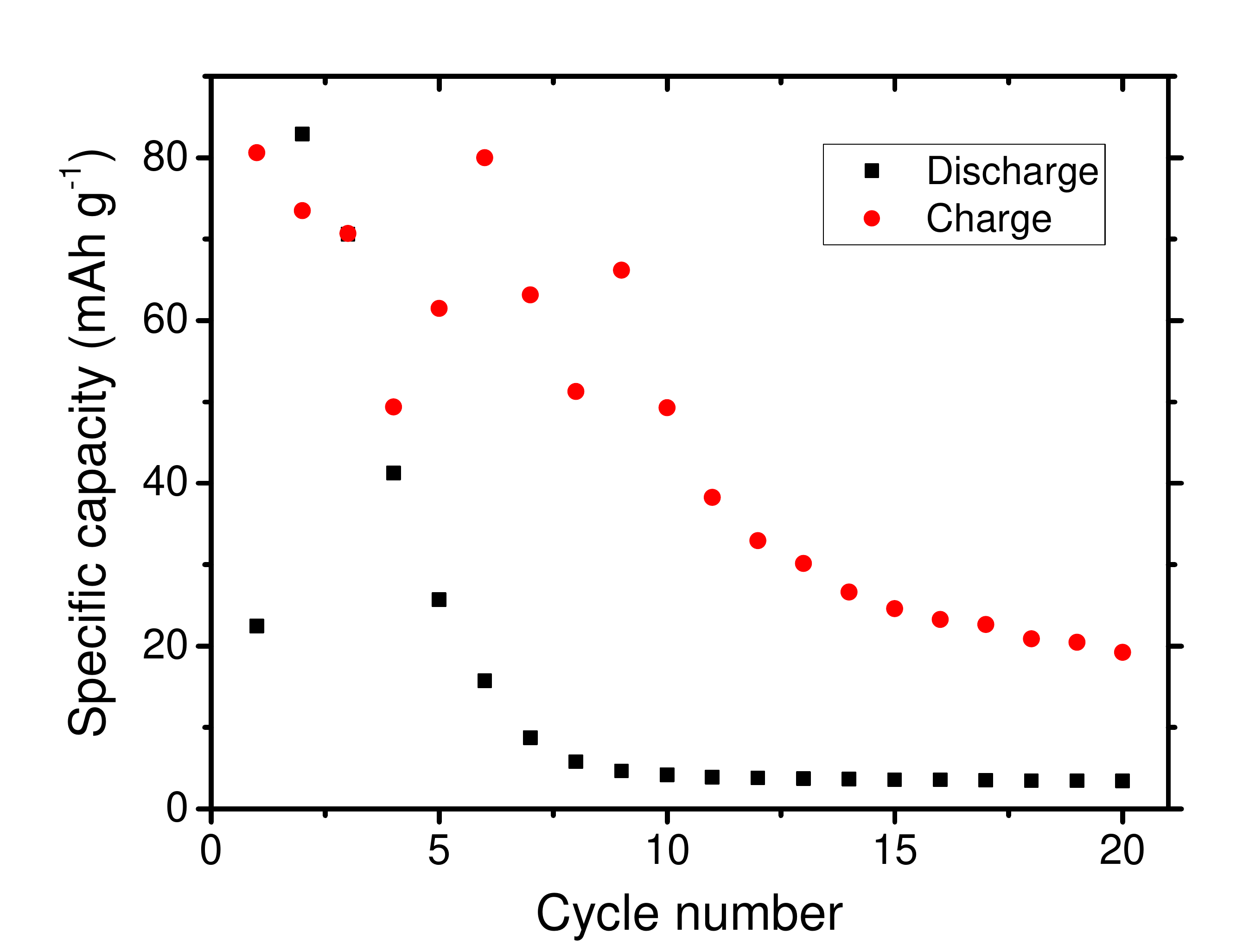}
	\includegraphics[width=0.49\textwidth]{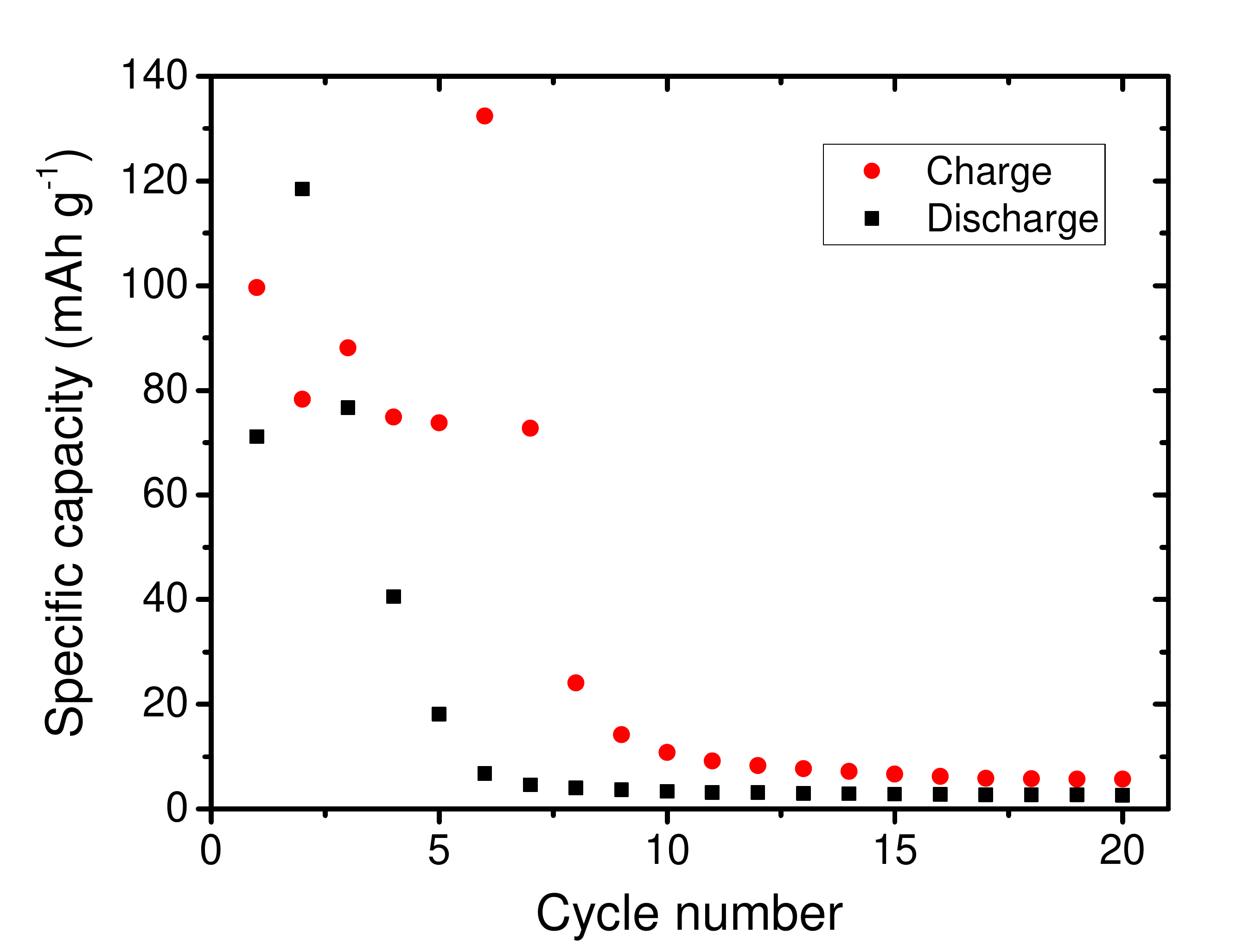}
	\caption{Specific capacities of typical galvanostatic charge-discharge tests (left: discharge first, right: charge charge first. Current rate: 100 mA g$^{-1}$) of a Swagelok-type cell built using core/sheath C/\ce{V2O5} nanofibres as cathode. The plots show that the capacity fades quickly in both experiments, due to the disintegration of the \ce{V2O5}}
	\label{fig:CoreSheathCapFade}
\end{figure}

\begin{figure}[h!]
	\centering
    	\includegraphics[width=0.45\textwidth]{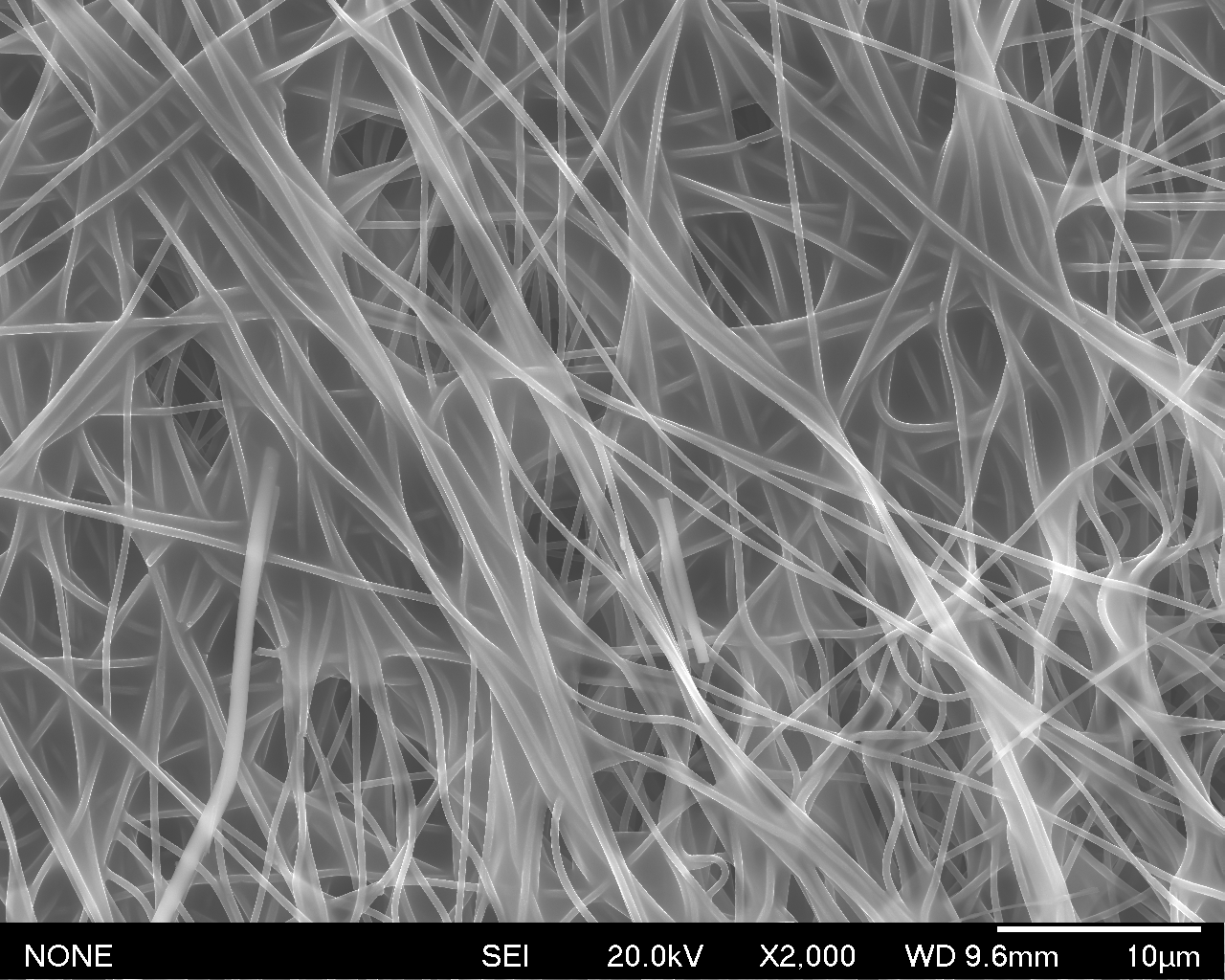}
	\includegraphics[width=0.5\textwidth]{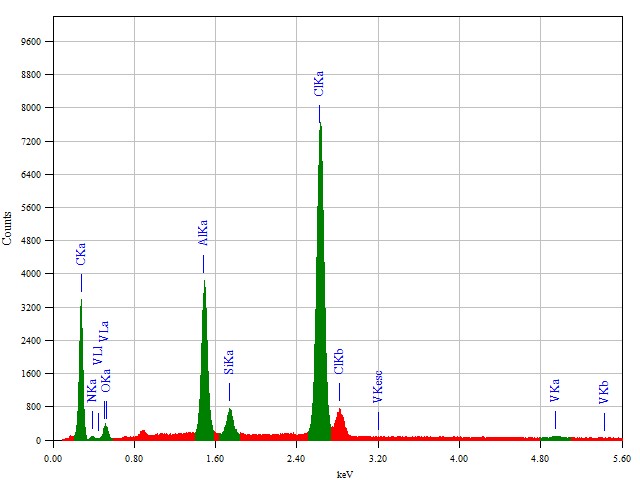}

	\caption{\textit{Ex situ} SEM (top) and EDX analysis (bottom) were performed on a cathode after running 20 cycles at 100 mA g$^{-1}$ to verify the hypothesis that the cycling process causes the disintegration of the \ce{V2O5} in the material. It can be seen  that the vanadium oxide sheath has been almost completely removed from the nanofibres, and they are now coated in an SEI film. The elemental composition of the film is very similar to what was observed in our previous work.\cite{canever_solid-electrolyte_2020} This confirms that the \ce{V2O5} sheath gets disintegrated in the charging-discharging process. The presence of a Si K$_\alpha$ peak in the EDX spectrum is likely imputable to some residual glass microfibre in the sample from the separator used to assemble the battery device.}
	\label{fig:CoreSheathCycledSEMEDX}
\end{figure}

\begin{figure}[h!]
	\centering
	\includegraphics[width=0.49\textwidth]{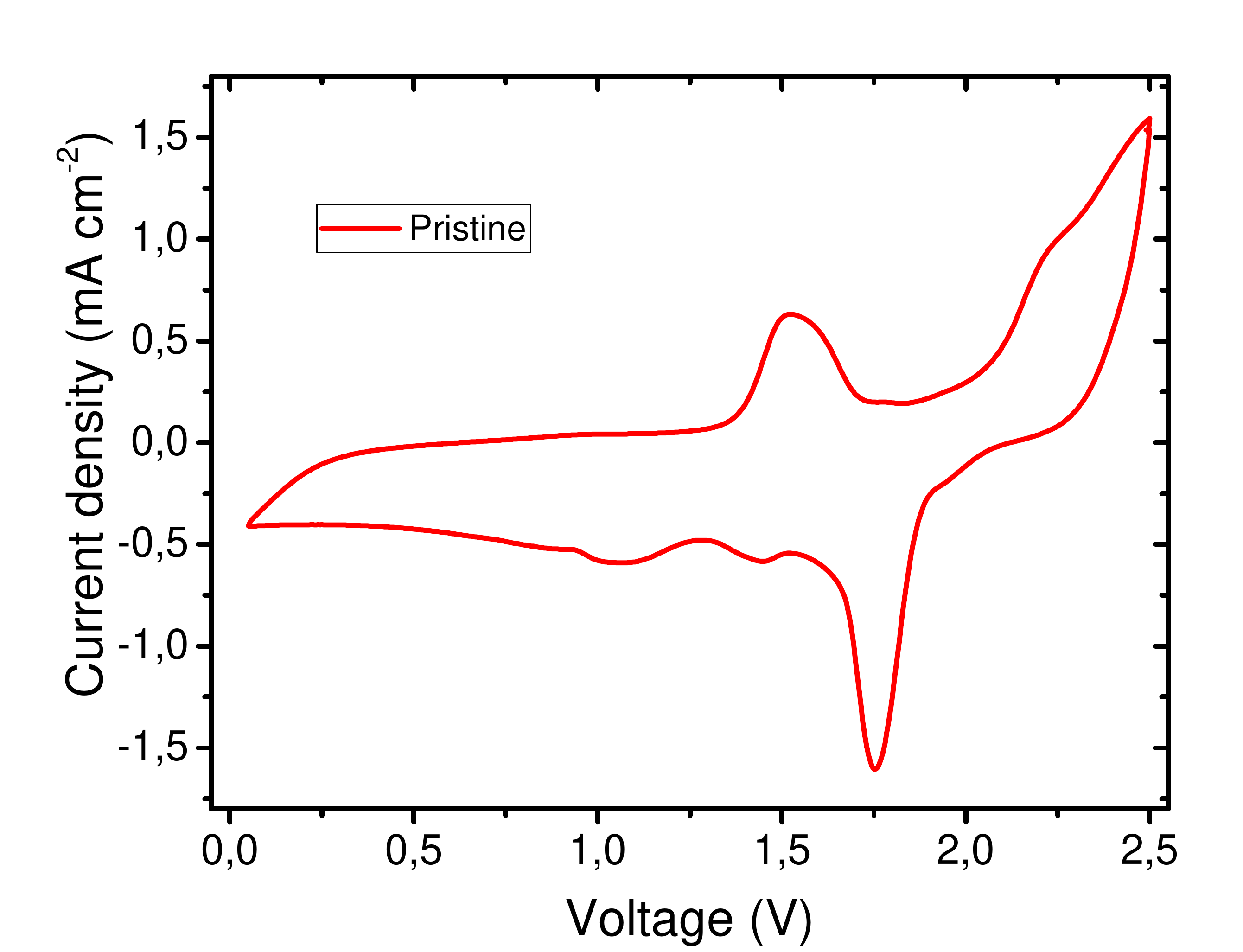}
	\includegraphics[width=0.49\textwidth]{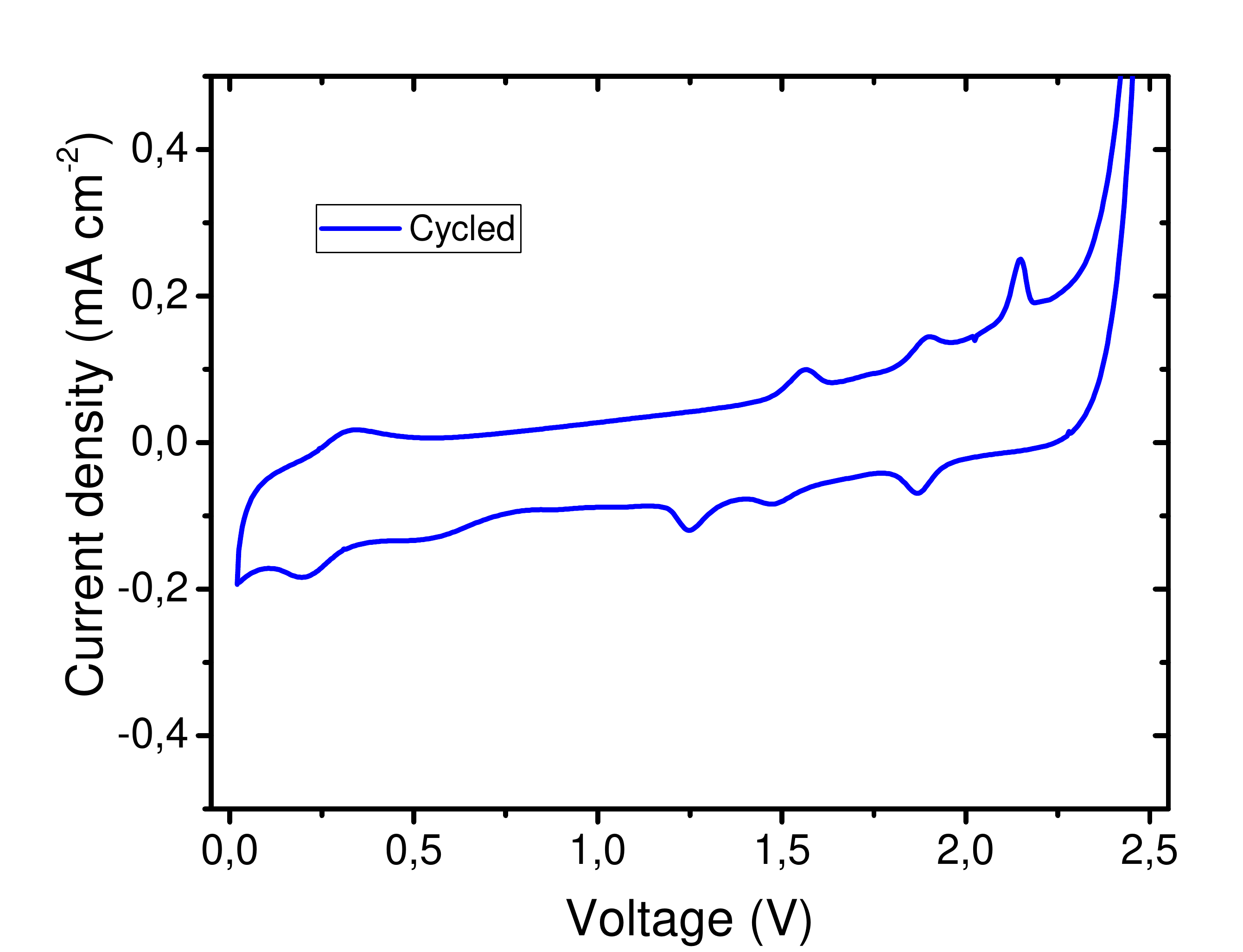}
	\caption{Cycling voltammograms of a device built using core/sheath C/\ce{V2O5} nanofibres, before (Left) and after (Right) performing 20 charge-discharge cycles at 100 mA g$^{-1}$. 
Further evidence for the disintegration of the \ce{V2O5} sheath can be found here: the CV performed on the freshly-assembled device shows an overall high current intensity, and a series of peaks 
imputable to the intercalation of \ce{Al^3+} and \ce{AlCl4-} ions. After galvanostatic cycling, however, the current density decreases notably, and the peaks are much less intense. A sharp increase in current beyond 2.25 V is also now present, suggesting that the carbon core is now likely exposed and promoting the formation of the SEI.
}
	\label{fig:CoreSheathCVbeforeafter}
\end{figure}

\begin{figure}[h!]
	\centering
	\includegraphics[width=0.49\textwidth]{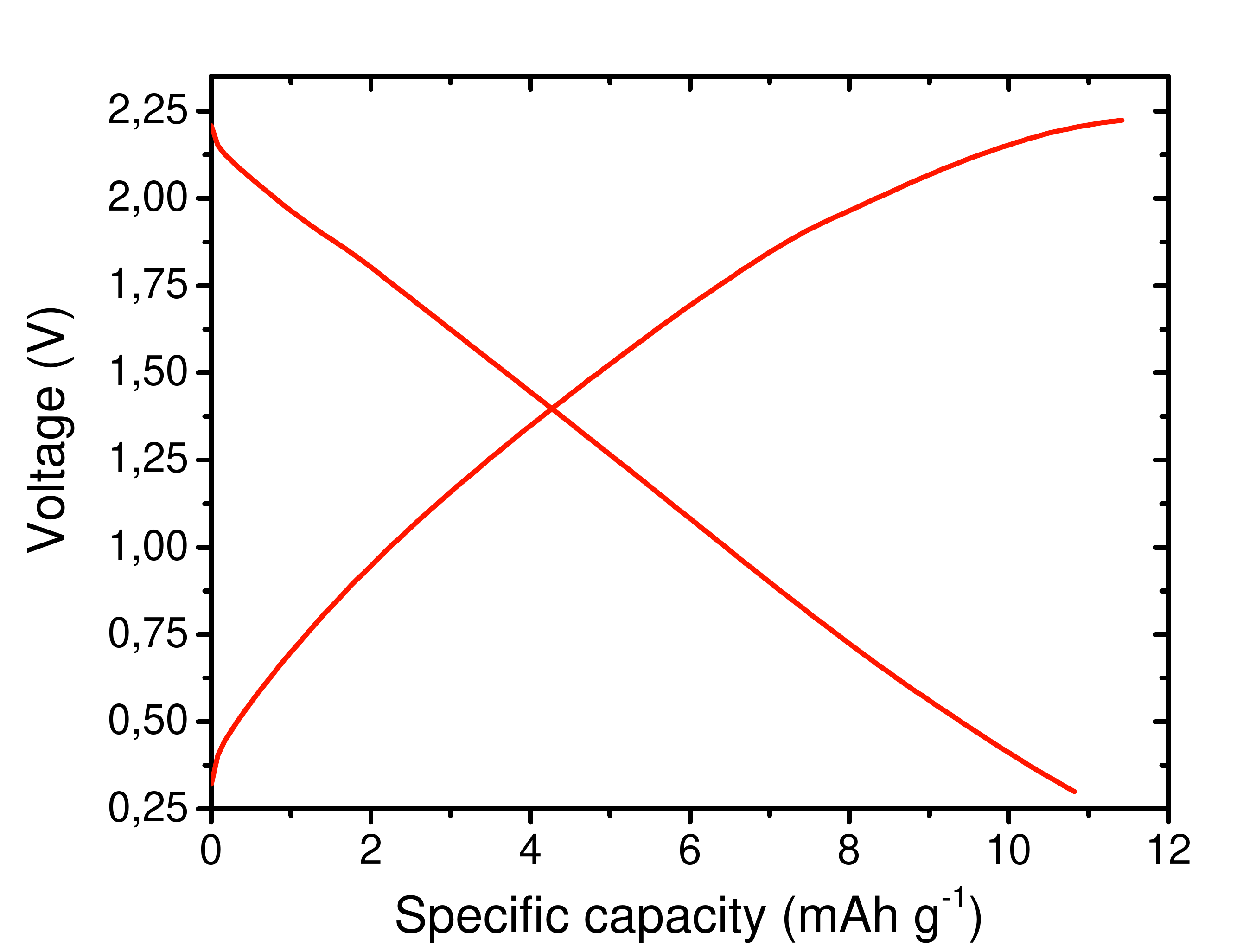}
	\includegraphics[width=0.49\textwidth]{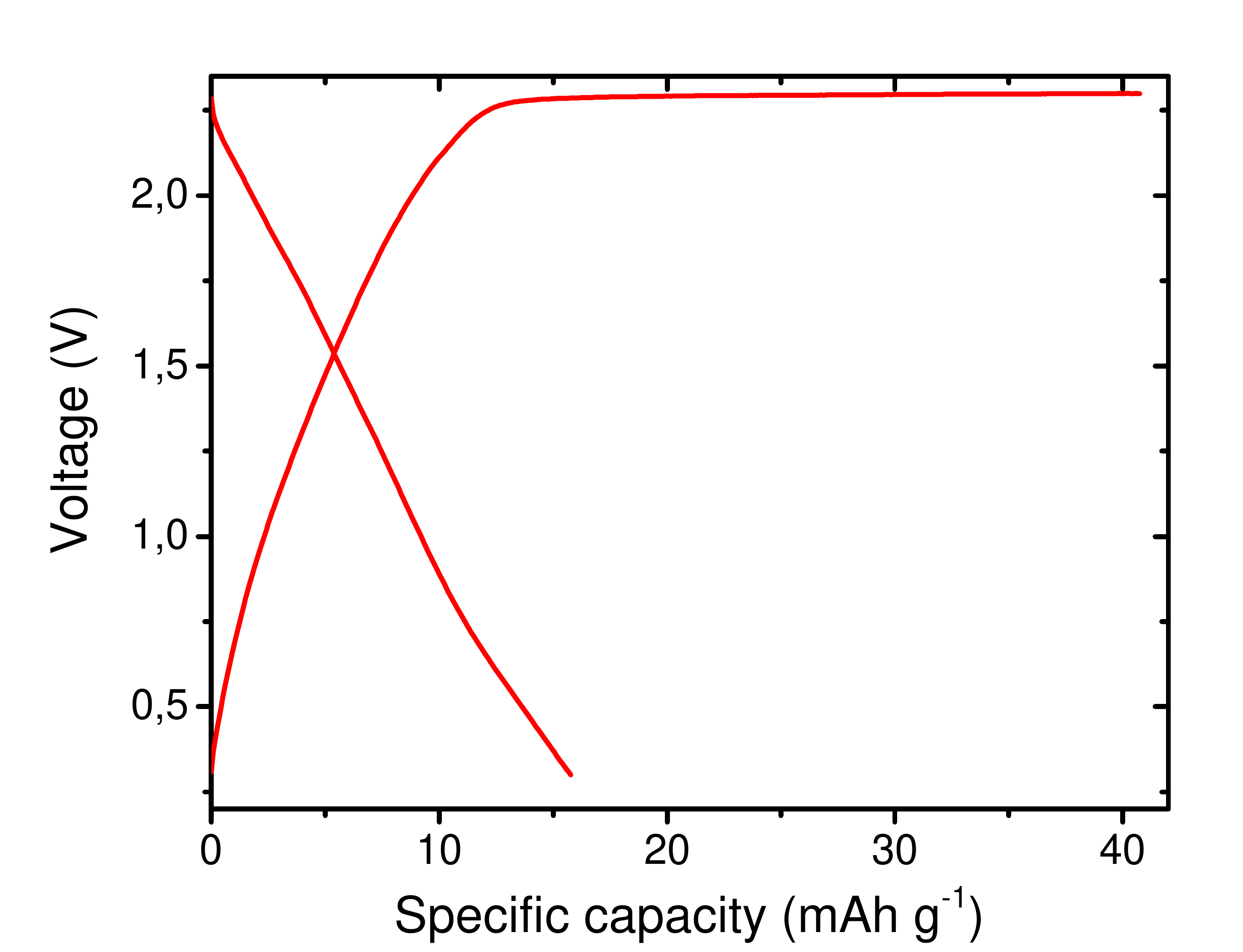}
		\caption{Galvanostatic charge-discharge profiles (fifth cycle, 100 mA g$^{-1}$) of a device built using pure carbon nanofibres, using a charging cut-off potential of 2.25 V (left) and 2.3 V (right). It can be seen at higher charging potentials the oxidative decomposition of the electrolyte occurs.\cite{canever_solid-electrolyte_2020}}
\label{PureCNFCD}
\end{figure}

\clearpage

\bibliographystyle{unsrt} 
\bibliography{refs}